\pdfoutput=1
\documentclass[sn-mathphys,Numbered]{sn-jnl}

\usepackage{graphicx}%
\usepackage{multirow}%
\usepackage{amsmath,amssymb,amsfonts}%
\usepackage{amsthm}%
\usepackage{mathrsfs}%
\usepackage[title]{appendix}%
\usepackage{xcolor}%
\usepackage{textcomp}%
\usepackage{manyfoot}%
\usepackage{booktabs}%
\usepackage{algorithm}%
\usepackage{algorithmicx}%
\usepackage{algpseudocode}%
\usepackage{listings}%
\usepackage{setspace}
\usepackage{lineno}
\usepackage{tikz}
\usepackage{float}
\usetikzlibrary{decorations.pathreplacing}
\usepackage{placeins}

\newcount\colveccount
\newcommand*\colvec[1]{
        \global\colveccount#1
        \begin{pmatrix}
        \colvecnext
}
\def\colvecnext#1{
        #1
        \global\advance\colveccount-1
        \ifnum\colveccount>0
                \\
                \expandafter\colvecnext
        \else
                \end{pmatrix}
        \fi
}

\theoremstyle{thmstyleone}%
\theoremstyle{thmstyletwo}%

\theoremstyle{thmstylethree}%

\begin{document}

\title[Article Title]{Operator Space Manifold Theory: Modeling Quantum Operators with a Riemannian Manifold}


\author*[]{\fnm{Gabriel} \sur{Nowaskie}}\email{gabriel.nowaskie341@topper.wku.edu}

\affil*[]{\orgdiv{Department of Physics and Astronomy}, \orgname{The Gatton Academy, Western Kentucky University}, \orgaddress{\street{1906 College Heights Blvd.}, \city{Bowling Green}, \postcode{42101}, \state{KY}, \country{United States}}}


\abstract{The Half-Transform Ansatz (HTA) is a proposed method to solve hyper-geometric equations in Quantum Phase Space by transforming a differential operator to an algebraic variable and including a specific exponential factor in the wave function, but the mechanism which provides this solution scheme is not known. Analysis of the HTA's application to the Hydrogen atom suggests an underlying mechanism which the HTA is a part of. Observations of exponential factors that act on the wave function naturally suggest modeling quantum operator definitions as a point on a Riemannian manifold in the 4D Operator Space, a novel idea we call the Operator Space Manifold Theory. Expanding on this concept, we find the true nature of the HTA and how Operator Space Manifold Theory can be used to describe and solve quantum systems by manipulating how a quantum state perceives position and momentum.}




\maketitle

\section{Introduction}\label{sec1}

With the discovery of the QPSR by Torres-Vega and Frederick [1], a framework with the capabilities to describe quantum systems in phase space, and the Heaviside Operational Ansatz (HOA) was created and published in 2004 by Simpao to generate exact analytical wave functions in the QPSR [2, 3]. The HOA yields implicit solutions in Quantum Phase Space. By applying a Laplace transform on the $p$ differential operator, we developed a general scheme that seemed to work for finding explicit solutions to the Schrodinger Equation or any similar hyper-geometric equation with the use of the Nikiforov-Uvarov (NU) Method [4]. This was called the Half-Transform Ansatz (HTA), but there is a lack of theoretical development justifying this method or explaining its nature.\\

 This work aims to produce a theoretical backing to the HTA as continuation to historical developments in [4], where we develop our Operator Space Manifold Theory. We develop this theory in the following sections and this paper is organized as follows. Section 2 is a recap of the Quantum Phase Space formulation and the HOA. Section 3 is a brief summary of the NU method. Section 4 provides the original HTA as in [4]. Section 5 shows the application of the HTA to the Hydrogen atom and its discrepancies with the configuration space results. Section 6 presents the Operator Space Manifold Theory. Section 7 presents how to traverse the Operator Space Manifold by the use of transforming one of the differential operators to an algebraic variable. Section 7 concludes our results and presents a full description of the HTA with the insight gained from the Operator Space Manifold Theory.

\bigskip

\section{Recap of the QPSR and Hamiltonian Construction}
\label{section:section2}

To begin, we must revisit the Heaviside Operational Ansatz. Only the brief methods relevant to this work are presented. Historical developments can be seen in [2, 3]. The variable relations: $x,p,t$ are the configuration space position, momentum, and time respectively. When transforming to the QPSR we recall that (where \ $\hat{}$ \ denotes operators and $\alpha,\gamma$ are otherwise free parameters as in [1]): 
\begin{equation*}
    \begin{aligned}
        H\left(x,\ p,\ t\right)\rightarrow\ \hat{H}\left(\hat{x},\ \hat{p},\ t\right)=\hat{H}\left(i\hbar\partial_p+\alpha x,\ -i\hbar\partial_x+\gamma p,\ t\right),\ \ni\alpha+\gamma=1
    \end{aligned}
\end{equation*}
\begin{equation*}
    \begin{aligned}
       x\rightarrow\ \hat{x}\equiv\ i\hbar\partial_p+\alpha\ x,\ p\rightarrow\hat{p}\equiv -i\hbar\partial_x+\gamma\ p,t\rightarrow\ t=t 
    \end{aligned}
\end{equation*}
\begin{equation*}
    \begin{aligned}
       \left(x_1,\ldots,x_n\right)\rightarrow\left(\widehat{x_1},\ldots,\widehat{x_n}\right)= \left(i\hbar\partial_{p_1}+\alpha_1x_1,\ldots,i\hbar\partial_{p_n}+\alpha_nx_n\right),\ni\alpha_j+\gamma_j=1,\ j=1,\ldots,n
    \end{aligned}
\end{equation*}
\begin{equation*}
    \begin{aligned}
           H\left(x_1,\ldots,x_n;p_1,\ldots,p_n; t\right)\rightarrow\hat{H}\left({\hat{x} }_1,\ldots,{\hat{x}}_n;{\hat{p} }_1,\ldots,{\hat{p}}_n;t\right)
    \end{aligned}
\end{equation*}
\begin{equation}
\label{eq:1}
    \begin{aligned}
        \equiv \hat{H}\left(i\hbar\partial_{p_1}+\alpha_1x_1,\ldots,i\hbar\partial_{p_n}+ \alpha_nx_n;-i\hbar\partial_{x_1}+\gamma_1p_1,\ldots,-i\hbar\partial_{x_n}+\gamma_np_n;t\right).
    \end{aligned}
\end{equation}

For our nucleus-electron system, our frame of reference is the nucleus. We thus transform to a spherical coordinate system and separate the angular variables, where $L(L+1)$ is the angular coupling term, ${p_r}$ is the radial momentum, and $r$ is the radial distance from the electron to the nucleus. For a 1D system, the Hamiltonian is:
\begin{equation}
    \begin{aligned}
        \hat{H}=\frac{{\hat{{p_r}}}^2}{2m}+\hat{V}(r).
    \end{aligned}
\end{equation}
For a 3D system, the Hamiltonian is:
\begin{equation}
\label{eq:3}
    \begin{aligned}
        \hat{H}=\sum^{N}{\frac{-i\hbar{\hat{p}}_r}{m\hat{r}}+\frac{{\hat{p}}_r^2}{2m}+\frac{\hbar^2L(L+1)}{2m{\hat{r}}^2}}+\hat{V}(r).
    \end{aligned}
\end{equation}
Typically, $\hat{r}=-i\hbar\frac{\partial}{\partial {p_r}}+\alpha r,\  \hat{p}_r=-i\hbar\frac{\partial}{\partial r}+\gamma {p_r}  $ by convention of the QPSR. The more general form presented in [3] is:
\begin{equation}
\label{eq:4}
    \begin{aligned}
        \hat{r}=\alpha r+i\hbar\beta\frac{\partial}{\partial {p_r}},\ \hat{{p_r}}=\gamma {p_r}+i\hbar\delta\frac{\partial}{\partial r}\ ,
    \end{aligned}
\end{equation}
where $\hat{r},\hat{p}_r$ are bound by the condition $\beta\gamma-\alpha\delta=1$ in order to keep consistency of $\hat{r}$ and $\hat{p}_r$ with the Heisenberg uncertainty principle, $[\hat{r},\hat{p}_r]=1$. 

\bigskip

\section{Nikiforov-Uvarov Method}
The Nikiforov-Uvarov method is based on solving the hyper-geometric type second-order differential equation of form
\begin{equation}
\label{eq:5}
    \begin{aligned}
        {\ \Psi}^{\prime\prime}\left(s\right)+\frac{\widetilde{\tau}}{\sigma\left(s\right)}\Psi^\prime\left(s\right)+\frac{\ \widetilde{\sigma}\left(s\right)}{\sigma^2\left(s\right)}\Psi\left(s\right)=0
    \end{aligned}
\end{equation}
Where $\sigma(s)$ and $\widetilde{\sigma}\left(s\right)$ are at most, second degree polynomials, and $\widetilde{\tau}\left(s\right)$ is at most a first-degree polynomial. $\Psi\left(s\right)$ is a function of hyper-geometric type. The solution of \eqref{eq:5} takes the form 
\begin{equation}
\label{eq:6}
    \begin{aligned}
        \Psi\left(s\right)=\phi\left(s\right)y\left(s\right)
    \end{aligned}
\end{equation}
Substituting \eqref{eq:6} into \eqref{eq:5} yields
\begin{equation}
    \begin{aligned}
        \sigma\left(s\right)y^{\prime\prime}\left(s\right)+\tau\left(s\right)y^\prime\left(s\right)+\lambda\left(s\right)y\left(s\right)=0,
    \end{aligned}
\end{equation}
where $\phi\left(s\right)$ satisfies the following relation
\begin{equation}
\label{eq:8}
    \begin{aligned}
        \frac{\phi^\prime\left(s\right)}{\phi(s)}=\frac{\pi\left(s\right)}{\sigma(s)}.
    \end{aligned}
\end{equation}
and $y\left(s\right)$ is a hyper-geometric type function, whose polynomial solutions are obtained from Rodrigues' relation 
\begin{equation}
    \begin{aligned}
        \ y\left(s\right)=y_n\left(s\right)=\frac{B_n}{\rho\left(s\right)}\frac{d^n}{ds^n}\left[\sigma^n\left(s\right)\rho\left(s\right)\right].
    \end{aligned}
\end{equation}
where $B_n$ is the normalization constant and $\rho\left(s\right)$ is a weight function satisfying the equation 
\begin{equation}
\label{eq:10}
    \begin{aligned}
        \left[\sigma\left(s\right)\rho\left(s\right)\right]^\prime=\tau\left(s\right)\rho\left(s\right).
    \end{aligned}
\end{equation}
The function $\pi\left(s\right)$ is defined as
\begin{equation}
\label{eq:11}
    \begin{aligned}
        \pi\left(s\right)=\left(\frac{\sigma^\prime-\widetilde{\tau}}{2}\right)\pm\sqrt{\left(\frac{\sigma^\prime-\widetilde{\tau}}{2}\right)^2-\widetilde{\sigma}+K\sigma},
    \end{aligned}
\end{equation}
and $\lambda$ is defined as 
\begin{equation}
\label{eq:12}
    \begin{aligned}
        \lambda=K+\pi\prime.
    \end{aligned}
\end{equation}
The value of $K$ can be calculated under the condition that the square root in \eqref{eq:11} must be the square of a polynomial. Thus, the equation of eigenvalues can be given as:
\begin{equation}
\label{eq:13}
    \begin{aligned}
        \lambda=\lambda_n=-n\tau^\prime-\frac{n\left(n-1\right)}{2}\sigma^{\prime\prime},
    \end{aligned}
\end{equation}
where
\begin{equation}
\label{eq:14}
    \begin{aligned}
        \tau\left(s\right)=\widetilde{\tau}\left(s\right)+2\pi\left(s\right).
    \end{aligned}    
\end{equation}
\bigskip

\section{The Half-Transform Ansatz}
Here we present the HTA. Its original statement can be seen in [4]. Using the general QPSR definitions from \eqref{eq:1}, the time independent wave equation becomes	
\begin{equation}
\label{eq:15}
    \begin{aligned}
        \hat{H}\binom{i\hbar\partial_{p_{r_1}}+\alpha_1x_1,\ldots,i\hbar\partial_{p_{r_n}}+\alpha_nx_n;}{-i\hbar\partial_{x_1}+\gamma_1p_{r_1},\ldots,-i\hbar\partial_{x_n}+\gamma_np_{r_n}}\psi\left(x_1,\ldots,x_n;p_{r_1},\ldots,p_{r_n}\right)=E_n\psi(x_1,\ldots,x_n;p_{r_1},\ldots,p_{r_n}). 
    \end{aligned}
\end{equation}
Applying the alternative to the convolution and multi-variable inverse transform (3A), the convolution of ${p_r}$ between (4A) and relation (5A) yields
\begin{equation*}
    \begin{aligned}
        \left[L_{\binom{\left({\partial_{p_{r_1}},\ldots,\partial_{p_{r_n}}}\right)}{\rightarrow\left(p_{r_1},\ldots.,p_{r_n}\right)}}^{-1}\left[\hat{H}\binom{i\hbar\partial_{p_{r_1}}+\alpha_1x_1,\ldots,i\hbar\partial_{p_{r_n}}+\alpha_nx_n;}{-i\hbar\partial_{x_1}+\gamma_1p_{r_1},\ldots,-i\hbar\partial_{x_n}+\gamma_np_{r_n}}\right]\right] \\
        \ast \ \ \ \ \ \psi\left(x_1,\ldots,x_n;p_{r_1},\ldots,p_{r_n}\right)\\ 
    \end{aligned}
    \begin{tikzpicture}[baseline=24pt]
        \hspace*{-145.5pt}
        \draw [decorate,decoration={brace,amplitude=5pt}](.6,0.0) -- (-.2,0.0);  
    \end{tikzpicture}
    \begin{tikzpicture}[baseline=33.5pt]
        \hspace*{-190.5pt}
        \node at (0,0) {$\left(p_{r_1},\ldots,p_{r_n}\right)$};
    \end{tikzpicture}
\end{equation*}
\begin{equation}
\label{eq:16}
    \begin{aligned}
        \equiv \hat{H}\binom{i\hbar\partial_{p_{r_1}}+\alpha_1x_1,\ldots,i\hbar\partial_{p_{r_n}}+\alpha_nx_n;}{-i\hbar\partial_{x_1}+\gamma_1p_{r_1},\ldots,-i\hbar\partial_{x_n}+\gamma_np_{r_n}}\psi\left(x_1,\ldots,x_n;p_{r_1},\ldots,p_{r_n}\right).                        
    \end{aligned}
\end{equation}
Applying \eqref{eq:16}-\eqref{eq:15} with the convolution identities in (3A), \eqref{eq:16} is transformed into
\begin{equation}
    \begin{aligned}
       L_{\binom{\left({p_{r_1},\ldots,p_{r_n}}\right)} {\rightarrow\left({\overline{p}_r}_1,\ldots.,{\overline{p}_r}_n\right)}}\ &\left[ \hat{H}\binom{i\hbar\partial_{p_{r_1}}+\alpha_1x_1,\ldots,i\hbar\partial_{p_{r_n}}+\alpha_nx_n;}{-i\hbar\partial_{x_1}+\gamma_1p_{r_1},\ldots,-i\hbar\partial_{x_n}+\gamma_np_{r_n}}\psi\left(x_1,\ldots,x_n;p_{r_1},\ldots,p_{r_n}\right)\right]
        \\ & \equiv\ L_{\binom{\left({p_{r_1},\ldots,{p_r}}_n\right)}{\rightarrow\left({\overline{p}_r}_1,\ldots.,{\overline{p}_r}_n\right)}}
        \begin{bmatrix}
        L_{\binom{\left({\partial_{p_{r_1}},\ldots,\partial}_{p_{r_n}}\right)}{\rightarrow\left(p_{r_1},\ldots.,p_{r_n}\right)}}^{-1}\left[\hat{H}\binom{i\hbar\partial_{p_{r_1}}+\alpha_1x_1,\ldots,i\hbar\partial_{p_{r_n}}+\alpha_nx_n;}{-i\hbar\partial_{x_1}+\gamma_1p_{r_1},\ldots,-i\hbar\partial_{x_n}+\gamma_np_{r_n}}\right] \\
        \ast \ \ \ \ \ \ \ \ \psi\left(x_1,\ldots,x_n;p_{r_1},\ldots,p_{r_n}\right)
        \end{bmatrix}
        \\
        &=E_n\check{\psi}\left(x_1,\ldots,x_n;{\overline{p}_r}_1,\ldots,{\overline{p}_r}_n\right)
    \end{aligned}
    \begin{tikzpicture}[baseline=23.5pt]

        \hspace*{-200pt}
        \draw [decorate,decoration={brace,amplitude=5pt}](.6,0.0) -- (-.2,0.0);  
    \end{tikzpicture}
    \begin{tikzpicture}[baseline=33pt]
        \hspace*{-242.5pt}
        \node at (0,0) {$\left(p_{r_1},\ldots,p_{r_n}\right)$};
    \end{tikzpicture}
\end{equation}
\begin{equation}
    \begin{aligned}
        \equiv\hat{H}\binom{i\hbar{\overline{p}_r}_1+\alpha_1x_1,\ldots,i\hbar{\overline{p}_r}_n+\alpha_nx_n;}{-i\hbar\partial_{x_1}+\gamma_1p_{r_1},\ldots,-i\hbar\partial_{ x_n}+\gamma_np_{r_n}}\check{\psi}\left(x_1,\ldots,x_n;{\overline{p}_r}_1,\ldots,{\overline{p}_r}_n\right) ={E}_n\check{\psi}\left(x_1,\ldots,x_n;{\overline{p}_r}_1,\ldots,{\overline{p}_r}_n\right).
    \end{aligned}
\end{equation}
Thus, the wave function in phase space can be expressed as the inverse transform of the solution $\breve{\psi}\left(x_1,\ldots,x_n;{\overline{p}_r}_1,\ldots,{\overline{p}_r}_n;t\right)$. The radial Schrodinger equation using the 3D full Cornell potential Hamiltonian is
\begin{equation}
    \begin{aligned}
        \left(-\frac{i\hbar\hat{{p_r}}}{m\hat{r}}+\frac{{\hat{{p_r}}}^2}{2m}+\frac{\hbar^2 L\left(L+1\right)}{2m\hat{r}}+V(\hat{r})\right)\psi\left(r,\overline{p}_r\right)=E_n\psi\left(r,\overline{p}_r\right).
    \end{aligned}
\end{equation}
We expand the ansatz previously stated in [4] by stating that the formulation:
\begin{equation}
    \begin{aligned}
        \psi\left(r,{p_r}\right)=e^{\frac{i{p_r}r}{\hbar}\left(\frac{\gamma}{\delta}\right)}\ \Omega\left(r,{p_r}\right),
    \end{aligned}
\end{equation}
where $\delta$ and $\gamma$ are defined as the coefficients in the momentum operator, degenerates the Schrodinger Equation in Quantum Phase Space to a form similar to the configuration space. By making the substitution $A=\alpha r+i\hbar\beta \overline{p}_r$, the Schrodinger equation can be written as an ordinary differential equation (ODE). This ODE is hyper-geometric, and thus can be solved using the Nikiforov Uvarov method. To then get the solution in the domain of $\left(r,p_r\right)$, an inverse Laplace transform from $(\overline{p}_r \to p_r)$ is done. The effects of Heisenberg's uncertainty principle and the nature of the Quantum Phase Space can be seen using the HTA. To present this, we present the application of the HOA to the Hydrogen Atom.

\bigskip 
\section{Application of the HTA to Hydrogen}
In this section, we apply the HTA to the 3D TISE in Quantum Phase Space. 

Using the general QPSR definitions from \eqref{eq:1} along with the Coulomb potential, the radial, time independent wave equation becomes	

\begin{equation}
    \begin{aligned}
        \left(-\frac{i\hbar\hat{{p_r}}}{m\hat{r}}+\frac{{\hat{{p_r}}}^2}{2m}+\frac{\hbar^2L(L+1)}{2m\hat{r}}-\frac{ke^2}{\hat{r}} \right)\psi\left(r,{\overline{p}_r}\right)=E_n\psi\left(r,\overline{p}_r\right).
    \end{aligned}
\end{equation}
Using the ansatz $\psi\left(r,{p_r}\right)=e^{\frac{i{p_r}r}{\hbar}\left(\frac{\gamma}{\delta}\right)} \Omega\left(r,{p_r}\right)$ along with the general operator definitions as in \eqref{eq:4} and mapping $(\partial_{{p_r}}\rightarrow\overline{p}_r)$:
\begin{equation}
    \begin{aligned}
        \begin{pmatrix}
            -\frac{2 e^2 k \Omega \left(r,\overline{p}_r\right) e^{\frac{i \gamma  r p_r}{\delta  \hbar }}}{\alpha  r+i \beta  \overline{p}_r \hbar }+ \frac{L^2 \hbar ^2 \Omega \left(r,\overline{p}_r\right) e^{\frac{i \gamma  r p_r}{\delta  \hbar }}}{m (\alpha  r+i \beta  \overline{p}_r \hbar )^2} \\ +\frac{L \hbar ^2 \Omega \left(r,\overline{p}_r\right) e^{\frac{i \gamma  r p_r}{\delta  \hbar }}}{m (\alpha  r+i \beta  \overline{p}_r \hbar )^2}+ \frac{2 \delta  \hbar ^2 e^{\frac{i \gamma  r p_r}{\delta  \hbar }}}{m (\alpha  r+i \beta  \overline{p}_r \hbar )}\frac{\partial \Omega \left(r,\overline{p}_r\right)}{\partial r}-\frac{\delta ^2 \hbar ^2 e^{\frac{i \gamma  r p_r}{\delta  \hbar }}}{m}\frac{\partial^2 \Omega \left(r,\overline{p}_r\right)}{\partial r^2}
            \end{pmatrix}=
            2 E_n \Omega \left(r,\overline{p}_r\right) e^{\frac{i \gamma  r p_r}{\delta  \hbar }}
    \end{aligned}
\end{equation}
which becomes
\begin{equation}
\label{eq:23}
    \begin{aligned}
        \begin{pmatrix}
        -\frac{2 e^2 k \Omega \left(r,\overline{p}_r\right) }{\alpha  r+i \beta  \overline{p}_r \hbar }+ \frac{L^2 \hbar ^2 \Omega \left(r,\overline{p}_r\right) }{m (\alpha  r+i \beta  \overline{p}_r \hbar )^2} \\ +\frac{L \hbar ^2 \Omega \left(r,\overline{p}_r\right) }{m (\alpha  r+i \beta  \overline{p}_r \hbar )^2}+ \frac{2 \delta  \hbar ^2 }{m (\alpha  r+i \beta  \overline{p}_r \hbar )}\frac{\partial \Omega \left(r,\overline{p}_r\right)}{\partial r}-\frac{\delta ^2 \hbar ^2 }{m}\frac{\partial^2 \Omega \left(r,\overline{p}_r\right)}{\partial r^2}
            \end{pmatrix}=
            2 E_n \Omega \left(r,\overline{p}_r\right) 
    \end{aligned}
\end{equation}

 We proceed with the substitution $A=\alpha r+i\hbar\beta \overline{p}_r$. The chain rule yields
\begin{equation}
    \begin{aligned}
        \ \frac{\partial\psi\left(r,\overline{p}_r\right)}{\partial r}=\frac{\partial\Omega(A)}{\partial A}\frac{\partial A}{\partial r}=\alpha \frac{\partial\Omega(A)}{\partial r},\ \ \frac{\partial^2\Omega\left(r,\overline{p}_r\right)}{\partial r^2}=\frac{\partial^2\Omega(A)}{\partial A^2}\frac{\partial A^2}{\partial r^2}=\alpha^2 \frac{\partial^2\Omega(A)}{\partial r^2}.
    \end{aligned}
\end{equation}
\eqref{eq:23} transforms into
\begin{equation}
    \begin{aligned}
        \Omega (A) \left(\frac{L^2 \hbar ^2}{A^2 m}+\frac{L \hbar ^2}{A^2 m}-\frac{2 e^2 k}{A}\right)-2 E_n \Omega (A)+\frac{2 \alpha  \delta  \hbar ^2 \Omega '(A)}{A m}-\frac{\alpha ^2 \delta ^2 \hbar ^2 \Omega ''(A)}{m}=0
    \end{aligned}
\end{equation}

\begin{equation}
\label{eq:26}
    \begin{aligned}
        =\frac{ \left(\frac{2 A^2 E_n m}{\hbar ^2}+\frac{2 A e^2 k m}{\hbar ^2}-L (L+1)\right)}{ A^2 \alpha^2 \delta^2}\Omega (A)-\frac{2 \Omega '(A)}{ A \delta \alpha}+\Omega ''(A)=0
    \end{aligned}
\end{equation}
For convenience, we define three constants, $\omega,\zeta,\kappa$, to be 
\begin{equation}
    \begin{aligned}
        \omega\equiv L (L+1),\ \ \zeta\equiv\ \frac{2 e^2 k m}{\hbar ^2},\ \ \kappa\equiv-\frac{2 E_n m}{\hbar ^2}.
    \end{aligned}
\end{equation}
Now, we can rewrite \eqref{eq:26} as
\begin{equation}
\label{eq:28}
    \begin{aligned}
        \frac{ \left(-\omega +A \zeta -A^2 \kappa \right)}{\alpha ^2 A^2 \delta ^2}\Omega (A)-\frac{2 \Omega '(A)}{\alpha  A \delta }+\Omega ''(A)=0,
    \end{aligned}
\end{equation}
the exact form as in \eqref{eq:5} to apply the Nikiforov-Uvarov method. Thus, we let 
\begin{equation}
    \begin{aligned}
        \sigma\left(A\right)=-\alpha \delta A, \quad \widetilde{\sigma}\left(A\right)=-\omega+A\zeta-A^2\kappa, \quad
        \widetilde{\tau}\left(A\right)=2. 
    \end{aligned}
\end{equation}
From \eqref{eq:11}
\begin{equation}
    \begin{aligned}
        \pi\left(A\right)=\frac{1}{2} (-\alpha  \delta -2)+\pm \sqrt{\frac{1}{4} (-\alpha  \delta -2)^2+A^2 \kappa -A \zeta -\alpha  A \delta  K+\omega }
    \end{aligned}
\end{equation}
Since the polynomial under the radical must a square of a polynomial, the discriminant must equal zero.
\begin{equation}
    \begin{aligned}
        (-4 \zeta -4 \alpha  \delta  K)^2-4 (4 \kappa ) \left(\alpha ^2 \delta ^2+4 \alpha  \delta +4 \omega +4\right)=0
    \end{aligned}
\end{equation}
Solving for $K$,
\begin{equation}
    \begin{aligned}
        K= -\frac{\alpha  \delta  \zeta +| \alpha  \delta |  \sqrt{\kappa  \left((\alpha  \delta +2)^2+4 \omega \right)}}{\alpha ^2 \delta ^2}, \quad \frac{-\alpha  \delta  \zeta + | \alpha \delta |  \sqrt{\kappa  \left((\alpha  \delta +2)^2+4 \omega \right)}}{\alpha ^2 \delta ^2}.
    \end{aligned}
\end{equation}
In an attempt to get a result similar to the configuration space Schrodinger equation, we want the expression inside the square inside the radical to be a perfect square, creating the condition: $\alpha \delta +2=1$. Thus,

\begin{equation}
\label{eq:33}
    \begin{aligned}
        \alpha=\frac{-1}{\delta}, \quad \alpha=\frac{-3}{\delta}.
    \end{aligned}
\end{equation}
These options are still bound by Heisenberg's uncertainty principle, $\beta \gamma -\alpha \delta =1$. We realize that the option $\alpha=\frac{-1}{\delta}$ will result in the configuration space operator definitions, and so our Schrodinger equation solution would become the configuration space solution. Thus, due to the constraint of Heisenberg's uncertainty principle and keeping this equation in phase space, we chose the option $\alpha=\frac{-3}{\delta}$, which makes the condition:
\begin{equation}
    \begin{aligned}
        \beta \gamma =-2.
    \end{aligned}
\end{equation}
Due to the exponential factor substitution earlier, it can be seen that the ansatz, $\psi(r,p_r)=e^{\left(\frac{i p_r r \gamma}{\hbar \delta}\right)}\Omega(r,p_r)$, truly is a degeneration of the PDE to the configuration space. However, this degeneration is restricted by Heisenberg's uncertainty principle, and the factor of 3 in \eqref{eq:33} is what embeds the intrinsic uncertainty between the position and momentum in our phase space equation.   We chose the value of $K$ that results in a negative derivative of $\tau$. Thus, we chose the negative sign. Simplifying K, we have
\begin{equation}
    \begin{aligned}
        K=\frac{1}{3} \left(\zeta -\sqrt{4 \kappa  \omega +\kappa }\right).
    \end{aligned}
\end{equation} 
This results in
\begin{equation}
\label{eq:36}
    \begin{aligned}
        \pi\left(A\right)=\frac{1}{2}\pm \frac{1}{2} \left(2 \sqrt{\kappa } A-\sqrt{4 \omega +1}\right)
    \end{aligned}.
\end{equation}
The choice of the sign in \eqref{eq:36} must once again generate a negative first derivative in $\tau$; thus, we take the negative sign again:
\begin{equation}
\label{eq:37}
    \begin{aligned}
        \pi\left(A\right)=\frac{1}{2}- \frac{1}{2} \left(2 \sqrt{\kappa } A-\sqrt{4 \omega +1}\right)
    \end{aligned}.
\end{equation}
From \eqref{eq:8}, $\phi(A)$ can be found by the differential equation
\begin{equation}
    \begin{aligned}
        \frac{\phi^\prime\left(A\right)}{\phi(A)}=\frac{\pi\left(A\right)}{\sigma(A)}=-\frac{-A \sqrt{\kappa }+\frac{1}{2} \sqrt{4 \omega +1}+\frac{1}{2}}{\alpha  A \delta }=\frac{-A \sqrt{\kappa }+\frac{1}{2} \sqrt{4 \omega +1}+\frac{1}{2}}{3 A},
    \end{aligned}
\end{equation}
which yields the exponential solution
\begin{equation}
    \begin{aligned}
        \phi\left(A\right)=e^{\frac{A \sqrt{\kappa }}{\alpha  \delta }} A^{-\frac{\sqrt{4 \omega +1}+1}{2 \alpha  \delta }}=e^{-\frac{1}{3} \left(A \sqrt{\kappa }\right)} A^{\frac{1}{6} \left(\sqrt{4 \omega +1}+1\right)}.
    \end{aligned}
\end{equation}
By \eqref{eq:14}, $\tau$ is 
\begin{equation}
    \begin{aligned}
        \tau\left(A\right)=-2 A \sqrt{\kappa }+\sqrt{4 \omega +1}+3.
    \end{aligned}
\end{equation}
\eqref{eq:10} is used to find the weight function $\rho\left(A\right)$, where 
\begin{equation}
    \begin{aligned}
        \left[\sigma\left(A\right)\rho\left(A\right)\right]^\prime=\tau\left(A\right)\rho\left(A\right),
    \end{aligned}
\end{equation}
\begin{equation}
    \begin{aligned}
        \rho (A) \left(\alpha  \delta -2 A \sqrt{\kappa }+\sqrt{4 \omega +1}+3\right)+\alpha  A \delta  \rho '(A)=0,
    \end{aligned}
\end{equation}
which gives the solution
\begin{equation}
    \begin{aligned}
        \rho\left(A\right)=e^{\frac{2 A \sqrt{\kappa }}{\alpha  \delta }} A^{-\frac{\alpha  \delta +\sqrt{4 \omega +1}+3}{\alpha  \delta }}=e^{-\frac{1}{3} \left(2 A \sqrt{\kappa }\right)} A^{\frac{1}{3} \sqrt{4 \omega +1}}.
    \end{aligned}
\end{equation}
With $\rho(A)$ and $\sigma\left(A\right)$, the function $y$ can be formulated as
\begin{equation}
    \begin{aligned}
        y_n\left(A\right)=\frac{B}{\rho\left(A\right)}\frac{d^n}{dA^n}\left(\sigma^n\left(A\right)\rho\left(A\right)\right).
    \end{aligned}
\end{equation}
where $B$ is a normalization constant. Recalling \eqref{eq:6}, the wave function must be 
\begin{equation*}
    \begin{aligned}
        \psi_n\left(A\right)=e^{\frac{i{p_r}r}{\hbar}\left(\frac{\gamma}{\delta}\right)} \Omega(A)=e^{\frac{i{p_r}r}{\hbar}\left(\frac{\gamma}{\delta}\right)}y_n\left(A\right)\phi\left(A\right)=e^{\frac{i{p_r}r}{\hbar}\left(\frac{\gamma}{\delta}\right)}\phi\left(A\right)\frac{B}{\rho\left(A\right)}\ \frac{d^n}{dA^n}\left(\sigma^n\left(A\right)\rho\left(A\right)\right)
    \end{aligned}
\end{equation*}
\begin{equation}
\label{eq:45}
    \begin{aligned}
        =B e^{\frac{i{p_r}r}{\hbar}\left(\frac{\gamma}{\delta}\right)} e^{\frac{A \sqrt{\kappa }}{3}} A^{\frac{1}{6}-\frac{1}{6} \sqrt{4 \omega +1}} \frac{d^n}{dA^n}\left(e^{-\frac{1}{3} \left(2 A \sqrt{\kappa }\right)} A^{\frac{1}{3} \sqrt{4 \omega +1}} (-\alpha  A \delta )^n\right).
    \end{aligned}
\end{equation}
Thus, to find the wave function with respect to $(r,p_r)$, \eqref{eq:45} can be inversely Laplace transformed from $\overline{p_r}\to p_r$ after back-substituting $A$. Now, to cross check our results with the coordinate space hydrogen atom, we look at the energy eigenvalues using equations \eqref{eq:12} and \eqref{eq:13}:
\begin{equation}
    \begin{aligned}
        \lambda=K+\pi^\prime\left(A\right)= \frac{1}{3} \left(\zeta -\sqrt{4 \kappa  \omega +\kappa }\right)-\sqrt{\kappa },
    \end{aligned}
\end{equation}
\begin{equation}
    \begin{aligned}
        \lambda_n=\ -n\tau^\prime\left(A\right)-\frac{n\left(n-1\right)\sigma^{\prime\prime}\left(A\right)}{2}=2n \sqrt{\kappa }.
    \end{aligned}
\end{equation}
With condition $\lambda=\lambda_n$, 
\begin{equation}
\label{eq:48}
    \begin{aligned}
        \frac{1}{3} \left(\zeta -\sqrt{4 \kappa  \omega +\kappa }\right)-\sqrt{\kappa }=2 \sqrt{\kappa } n
    \end{aligned}
\end{equation}
Solving this equation for $\kappa$, we have
\begin{equation}
    \begin{aligned}
        \kappa =\frac{\zeta ^2}{\left(6 n+\sqrt{4 \omega +1}+3\right)^2}
    \end{aligned}
\end{equation}
Back-substituting $\omega\equiv L (L+1),\ \zeta\equiv\ \frac{2 e^2 k m}{\hbar ^2},\ \kappa\equiv-\frac{2 \text{En} m}{\hbar ^2}$, we find the energy eigenvalue equation to be 
\begin{equation}
\label{eq:50}
    \begin{aligned}
        E_n=-\frac{e^4 k^2 m}{2 \hbar ^2 (L+3 n+2)^2}.
    \end{aligned}
\end{equation}
This result is not the same as the configuration space solution, and is a direct consequence of the relationship $\alpha=\frac{-3}{\delta}.$ If we were to take \eqref{eq:48} and \eqref{eq:45} and have the variables $\alpha$ and $\delta$ not simplified, we would see that for $\alpha \to \frac{-1}{\delta}$ and $\gamma=0$ (which conforms with the momentum operator definition in the configuration space), these expressions would converge to the exact Hydrogen atom wave function and energy eigenvalues. Because we chose to have a wave function including both $p$ and $r$, the Heisenberg uncertainty principle embedded a factor of $3$ that represents the error in measuring $r$ and $p$ simultaneously. This error uniquely spread across the energy eigenvalues and the wave function depending on what path you take in solving \eqref{eq:28}. For the values of $\alpha, \beta, \gamma, \delta$ that define the coordinate space or the momentum space, the inverse Laplace transform isn't needed since the coefficient in front of $p$ or $r$ will then be $0$. Though, this requires a mechanism that exists which dictates the nature of the operator coefficients $\alpha, \beta, \gamma, \delta$ the Schrodinger Equation, or any quantum equation, before this inverse transform. In other words, there must then exist a framework that defines the nature of the quantum operator coefficients that is independent of the $\overline{p}$ space previously used. We also need this framework to describe the nature of the $e^{\frac{i{p_r}r}{\hbar}\frac{\gamma}{\delta}}$ factor.

\bigskip

\section{Operator Space Manifold Theory}
In this section we present a mechanism to describe all of the concerns in the HTA that generates a overarching theory to describe the Quantum Phase Space. \\

Consider a 4D space which has 4 degrees of freedom: $\alpha, \beta, \gamma, \delta.$ In this space there is a curve that is represented by $\beta \gamma -\alpha \delta =1$, such that Heisenberg's uncertainty principle lives on this curve and every point on this curve represents a possible definition of phase space. Looking at \eqref{eq:23}, $e^{\frac{i{p_r}r}{\hbar}\frac{\gamma}{\delta}}$ seems to have the the effect of forcing $\gamma$ to converge to $0$. We interpret this exponent as a rotation of the form $e^{i \phi}$ and assume that this exponential factor is an operator that translates a point on the curve $\beta \gamma -\alpha \delta =1$ to have $\gamma=0$; therefore, each point on our curve $\beta \gamma - \alpha \delta$ must have defined angles. This naturally leads us to use a Riemannian manifold as the machinery for our framework. \\

More formally stated, let us consider a Riemannian manifold defined by $\beta \gamma - \alpha \delta=1$ in the $\alpha, \beta, \gamma, \delta$ 4D Operator Space. A "specialized" phase space is the space with operator definitions defined by any point on the manifold; we can think of these specialized phase spaces as the image of a point on the manifold. We consider the "entire" phase space to be the complete set of specialized phase spaces, or the set of all images from all points on the manifold. By definition, this manifold also includes the coordinate space and the momentum space operator definitions, which are seen as subsets of the entire phase space.\\

 We define each point in the Operator Space by:

\begin{equation*}
    \colvec{4}{\alpha}{\beta}{\gamma}{\delta}.
\end{equation*}
 
 $e^{\frac{i{p_r}r}{\hbar}\frac{\gamma}{\delta}}$ is seen as an operator that converges $\gamma \to 0$, and thus is defined to be a rotational transform on this manifold such that angle $\phi = \frac{{p_r}r}{\hbar}\frac{\gamma}{\delta}$ is the angle between the tangent space of a particular point and the 3D space where $\gamma$ is one less than initially ($1$ in this scenario). To mathematically represent this we postulate that 
 \begin{equation}
 \label{eq:51}
     \begin{aligned}
         e^{i \phi}\psi = g_\eta \colvec{4}{\alpha}{\beta}{\gamma}{\delta}
     \end{aligned}
 \end{equation}
where $g_\eta$ is a $4 \times 4$ matrix that acts as the linear transform in Operator Space analog to a phase shift by $\phi$. These rotational transforms can also be seen as a translation to a specific point $(\alpha, \beta, \gamma, \delta)$. For simplicity, we only consider this translational definition continuing forward. \eqref {eq:51} is also a mathematical way to represent the idea that a quantum system is analog to a point on the Operator Space Manifold. The effects of a specific $\phi$ on the Operator Space can be seen when applying $e^{i \phi}$ onto $\psi$. We present a table of the most basic transformations:

\begin{table}[ht]

\centering
\caption{4 fundamental $g_{\eta}$ matrices for different $\phi$.\label{tabel:1}}
\smallskip
\begin{tabular}{@{}c|c@{}}
\toprule
$\phi$ & $g_\eta$\\
\midrule
\\
\(\displaystyle \phi_1 = \frac{p_r r}{\hbar} \left(\frac{\alpha}{\beta}\right) \) & \(\displaystyle \begin{pmatrix}
    0& 0& 0& 0& \\
    0& 1& 0& 0& \\
    0& 0& 1& 0& \\
    0& 0& 0& 1& \\
\end{pmatrix}={g_\eta}_1\) \\
\\
\hline
\\
\(\displaystyle  \phi_3 = \frac{p_r r}{\hbar} \left(\frac{\gamma}{\delta}\right) \) & \(\displaystyle \begin{pmatrix}
    1& 0& 0& 0& \\
    0& 1& 0& 0& \\
    0& 0& 0& 0& \\
    0& 0& 0& 1& \\
\end{pmatrix}={g_\eta}_3\) \\
\\
\hline
\\
\(\displaystyle  \phi_2 = i \log \left(\psi \left(r,p_r\right)\right)+C\times r \) & \(\displaystyle \begin{pmatrix}
    1& 0& 0& 0& \\
    0& 0& 0& 0& \\
    0& 0& 1& 0& \\
    0& 0& 0& 1& \\
\end{pmatrix}={g_\eta}_2\) \\
\\
\hline
\\
\(\displaystyle  \phi_4 = i \log \left(\psi \left(r,p_r\right)\right)+C\times p_r \) & \(\displaystyle \begin{pmatrix}
    1& 0& 0& 0& \\
    0& 1& 0& 0& \\
    0& 0& 1& 0& \\
    0& 0& 0& 0& \\
\end{pmatrix}={g_\eta}_4\) \\
\botrule
\end{tabular}
\end{table}

\FloatBarrier

where $C$ is an arbitrary complex constant. The indices $n$ for ${g_\eta}_n$ are simply decided by which $1$ across the diagonal becomes $0$. For notation simplicity when developing the algebra of these matrices, we also introduce the idea of the a complement matrix, ${g_\eta}_C$:
\begin{equation}
    \begin{aligned}
        {g_\eta}_{C} = I -{g_\eta}
    \end{aligned}
\end{equation}
where $I$ is the identity matrix. The complements of the $g_\eta$ matrices (${g_\eta}_C$) represent the manipulation on a specific operator coefficient, or the causality, while the $g_\eta$ matrices represent the result of that manipulation on the entire space, or the effect. \\

When applying  $e^{\frac{i{p_r}r}{\hbar}\frac{\gamma}{\delta}}$ twice to $\psi(r,p_r)$ we have
\begin{equation}
\label{eq:53}
    \begin{aligned}
        i \delta  \hbar  \psi ^{(1,0)}\left(r,p_r\right) e^{\frac{2 i \gamma  r p_r}{\delta  \hbar }}-\gamma  p_r \psi \left(r,p_r\right) e^{\frac{2 i \gamma  r p_r}{\delta  \hbar }},
    \end{aligned}
\end{equation}
a result corresponding to
\begin{equation}
    \begin{aligned}
        g_\eta=\begin{pmatrix}
            1& 0& 0& 0& \\
            0& 1& 0& 0& \\
            0& 0& -1& 0& \\
            0& 0& 0& 1& \\
        \end{pmatrix}.
    \end{aligned}
\end{equation}
Applying this transform $n$ times, we see a pattern: each transform shifts the index on the diagonal affecting $\gamma$ by $-1$ each time, starting originally from the identity matrix $I$. In other words, the original $g_\eta$ matrix is shifted by ${g_\eta}_C$ n times, where n is the number of times you apply the specific operator. This same result holds true for the other transforms and their respective operator coefficients they manipulate, and we present a mathematical representation of this below: 
\begin{equation}
\label{eq:55}
    \begin{aligned}
        {g'_\eta} = - n {g_\eta}_C + {g_0},
    \end{aligned}
\end{equation}
where $g'_\eta$ is the resultant $g_\eta$ vector, ${g_0}$ is the initial $g_\eta$ matrix you are working with (typically the Identity matrix $I$ to represent the entire Operator Space), and ${g_\eta}_C$ is the complement of the $g_\eta$ matrix for the specific operator you are applying to $\psi$. For example, if we were to begin with the entire Operator Space ($g_0 = I$), and applied $e^{\frac{i{p_r}r}{\hbar}\frac{\gamma}{\delta}}$ 2 times, we would have the resultant $g'_\eta$ matrix:
\begin{equation}
    \begin{aligned}
        g'_\eta = -2 \begin{pmatrix}
            0& 0& 0& 0& \\
            0& 0& 0& 0& \\
            0& 0& 1& 0& \\
            0& 0& 0& 0&
        \end{pmatrix} + \begin{pmatrix}
            1& 0& 0& 0& \\
            0& 1& 0& 0& \\
            0& 0& 1& 0& \\
            0& 0& 0& 1&
        \end{pmatrix} = \begin{pmatrix}
            1& 0& 0& 0& \\
            0& 1& 0& 0& \\
            0& 0& -1& 0& \\
            0& 0& 0& 1&
        \end{pmatrix},
    \end{aligned}
\end{equation}
the same result as shown in \eqref{eq:53}. Thus, ${g_\eta}_1, {g_\eta}_2, {g_\eta}_3, {g_\eta}_4$ are unit transforms that shift $\alpha, \beta, \gamma, $ or $\delta$ by -1. We define the inverse of these $e^{i \phi}$ transforms to be when you make the substitution $\phi \to -\phi$, which then results in a shift of $+1$ instead of $-1$, or ${g_\eta}_C=-{g_\eta}_C$. \\

After defining these transforms that allow us to traverse the Operator Space Manifold, it is tempting to assume that they all must commute since the basis created from ${g_\eta}_1, {g_\eta}_2, {g_\eta}_3, {g_\eta}_4$ covers the entire Operator Space. However, these matrices have a restriction that appears due to the interaction of $\phi$ with differential operators creating an additional cross-term. Thus, we can only apply operators manipulating $\alpha$ and $\beta$ (${g_\eta}_1$ and ${g_\eta}_2$) or $\gamma$ and $\delta$ (${g_\eta}_3$ and ${g_\eta}_4$) simultaneously. The operators that can be applied simultaneously also commute. Therefore, we can only traverse the Operator Space by the position or the momentum operator coefficients at a time. This seems to be an expression of Heisenberg's uncertainty principle in the Operator Space. \\

This result, Heisenberg's uncertainty principle, cannot be worked around, as assuming there is a translational operator that is able to traverse the entire Operator Space at once, a cross term appears after the application of either the momentum or position operator which cannot be canceled due to the condition resolving to $e^{ix}=0$.\\

While we have stated that we cannot apply operators that will affect $\alpha, \beta, \gamma$, and $\delta$ simultaneously, what about these operators for equations or formulations that include both the momentum and position operators? Up until now, every formulation presented has been from applying single operators to $\psi(r,p_r)$ or choosing which operators can be applied. But the momentum and position operators will exist in almost every phase space formulation. For these scenarios, any of these operators on the Operator Space Manifold would inherently affect the entire Operator Space since its being applied to both the position and momentum operators. We will address this problem in the next section, which will also explain the importance of the HTA.\\

When applying multiple operators that are allowed to be used at the same time (those affecting only either the position or momentum operator coefficients at the same time), it is seen that the same principle from \eqref{eq:55} follows, except ${g_\eta}_{C}$ can is the addition of any of the ${g_\eta}_{C}$ from the operators being applied:
\begin{equation}
\label{eq:57}
    \begin{aligned}
        {g'_\eta} = -{g_\eta}_{C1} -{g_\eta}_{C2} -....- {g_\eta}_{Cn} + {g_0},
    \end{aligned}
\end{equation}
where ${g_\eta}_{Cn}$ is simply the ${g_\eta}_{C}$ for the nth operator that is being applied simultaneously. \\

We can summarize the algebra of these matrices as follows: only operators that act on $\alpha$ and $\beta$ or $\gamma$ and $\delta$ can be used simultaneously. The operators that can be used simultaneously commute, and a resultant ${g_\eta}_C$ matrix is found by simply summing the matrices. This same result holds true for applying the same operators multiple times. Thus, the Operator Space Manifold can only be simultaneously traversed using operators on either $\alpha$ and $\beta$ or $\gamma$ and $\delta$. The complement matrix represents the causality of operators, and can be seen as the shift from an original set of operator coefficient definitions to a new one.  This can be extended to any quantum system by stating that a system can only traverse space exactly by either $r$ or $p$. This result, known as Heisenberg's uncertainty principle, can thus find origins in the Operator Space. \\

We also briefly introduce the idea of "position-like" or "configuration-like" phase spaces and "momentum-like" phase spaces. We define "position-like" phase spaces to be specialized phase spaces with a $g_\eta$ of 
\begin{equation}
    \begin{aligned}
        g_\eta = \begin{pmatrix}
            1& 0& 0& 0& \\
            0& 0& 0& 0& \\
            0& 0& 0& 0& \\
            0& 0& 0& 1&
        \end{pmatrix},
    \end{aligned}
\end{equation}
and "momentum-like" phase spaces are specialized phase spaces with a $g_\eta$ of 
\begin{equation}
    \begin{aligned}
        g_\eta = \begin{pmatrix}
            0& 0& 0& 0& \\
            0& 1& 0& 0& \\
            0& 0& 1& 0& \\
            0& 0& 0& 0&
        \end{pmatrix}.
    \end{aligned}
\end{equation}
These two 2D spaces represent the analysis of a quantum system by its position or momentum. It is worth noting that the position and momentum space are subsets of these spaces respectively, and represent the most simple solutions to $\beta \gamma - \alpha \delta =1.$ This is possibly a representation of the Universe taking "least action" that is reflected within the  Operator Space Manifold.

\bigskip

\section{Traversing the Operator Space for Formulations with both Position and Momentum Operators}
In this section we will address the problem of traversing the Operator Space considering the restrictions due to Heisenberg's uncertainty principle suggest that it is not allowed. \\

By the principles of the Operator Space from the last section, we know that the Operator Space Manifold can only be traversed by operators affecting $\alpha, \beta$ or $\gamma, \delta$ simultaneously because the the cross-interaction of the differential operators and the operator coefficients. We also expressed how for equations or formulations that include both the momentum and position operators, any of the operators presented would inherently affect the entire Operator Space since its being applied to both the position and momentum operators. So, how are we supposed to traverse the Operator Space Manifold at all? \\

In order to bypass this restriction, we can take the Laplace transform one of the differential operators to an algebraic variable. This will redefine one of our quantum operators such that it will act like a variable. This move is the vital stepping stone of the HTA, and allows us to traverse the Operator Space Manifold by either $\alpha, \beta$ or $\gamma, \delta$ simultaneously. The restriction on only being able to traverse the Operator Space by either $\alpha, \beta$ or $\gamma, \delta$ still exists because of the cross-interaction between the angle $\phi$ in $e^{i \phi}$ and the derivative of the non-transformed quantum operator. Then, after solving the equation or doing the calculations, the inverse Laplace transform can be used to recover the information about $p$ or $r$. \\

But, if we apply a transform that makes either $\alpha,\gamma,\beta,\gamma$ become $0$ for the equation, then the inverse Laplace transform will simply return the original input. This is because the half transform used in the HTA can be changed to include the operator coefficient when taking the transform using the same logic in Section 4. If the coefficient operator is transformed to become 0, then the transform will be the input multiplied by the Dirac Delta function of $0$ which will return back the original input is not a function of either $\alpha r, \gamma p, \delta \overline{r},$ or $ \beta \overline{p}$ anymore. This fully explains how to recover the Hydrogen atom configuration space results from the phase space Schrodinger Equation, and can be used to recover any momentum or configuration space result from its phase space formulation.\\

The last thing to cover is the residual exponentials which are left behind after applying one of these Operator Space transformations such as in \eqref{eq:45} during the final formulation of the wave function. When applying these Operator Space transformations on a state, it is changing the way that state perceives its quantum operators. When such an ansatz of $\psi = e^{i \phi} \Omega$ is done in an equation, it is modeling that equation by assuming that there is a function $\Omega$ that is undergoing a transform from one point on the Operator Space Manifold to another. This changes how the equation $\psi$ is in will perceive position and momentum. \\

Because this transform is happening within $\psi$ and not on $\psi$, it does not change the way $\psi$ itself perceives position and momentum. Thus, while the equation of $\Omega$ perceives some transform in of ($\alpha, \beta, \gamma, \delta$), $\psi$ should not, and should still contain information about the entire Operator Space ($\alpha, \beta, \gamma, \delta$). \\ 

This conservation of information can be seen in the residual exponential factor when doing the inverse Laplace transform at the end of evaluating $\psi$, in which the residual exponential should factor out of the transform and still exist in the final wave function result. In summary, by using the HTA with the Operator Space Manifold, we can model a state, $\psi$, in equations by an Operator Space transform acting on a function, $\omega$, such that that $\omega$ has specific quantum operator definitions. These definition are isolated to only affect $\omega$ without actually losing that information in $\psi$. This information on the transformed operator coefficients is held within in these residual exponentials.  \\

\bigskip

\section{Conclusion: Operator Space Manifold Theory and the Nature of the HTA}
By applying the HTA as in [4] to the hydrogen atom and considering the differences between the phase space and the configuration space results, it is seen that there must be an underlying mechanism dictating the nature of $\alpha, \beta, \gamma$, and $\delta$ such that these values can be transformed to new values. Through observation of specific factor acting on $\psi$ and interpreting them as translations on some space, we develop the Operator Space Manifold Theory, which models each possible definition of the quantum operators as a point on a Riemannian manifold with specific restriction and behaviors as discussed in Sections 6 and 7.\\

From section 7, we know that the Operator Space Manifold cannot be traversed for equations with $\hat{r}$ and $\hat{p}$ unless a differential operator is transformed into an algebraic variable and that corresponding operator is represented as a collective variable. Thus, we present the HTA as: transforming a differential operator of a phase space formulation and then treating the corresponding quantum operator whose differential operator was transformed as a collective variable. An example for this method can be seen when. As posited in section 7, the HTA by definition is the method that allows us to traverse the Operator Space Manifold for equations that have $\hat{r}$ and $\hat{p}$. \\

Equations with $\hat{r}$ and $\hat{p}$ constitutes almost every formulation in phase space, and thus the HTA is the primary method to utilize the Operator Space Manifold to solve equations or generate Phase Space formulations. For equations or formulations that do not include both quantum operators, the HTA is not needed to traverse the Operator Space Manifold. However, both scenarios are still restricted by only traversing the Operator Space with transformations on only $\alpha, \beta$ or $\gamma, \delta$ simultaneously. \\

Traversing the Operator Space Manifold can be used to uniquely express equations and solve equations. One of these methods is the ansatz $\psi = e^{i \phi} \Omega$, where $e^{i \phi}$ is an Operator Space transformation. The simplest scenario is using one of the four fundamental transformations as in table \ref{tabel:1}. These unit transforms translate a point on the Operator Space manifold. When stating $\psi = e^{i \phi} \Omega$, the Operator Space transform $e^{i \phi}$ only acts on $\Omega$. In short, we can use this ansatz to model $\psi$ by a function which resides in a specific part of the Operator Space Manifold (has restricted or specific quantum operator definitions). Thus, $\Omega$ is perceiving its state by a different metric than $\psi$, but the information changed in $\Omega$ is only localized and is not reflected in $\psi$ because the residual factor $e^{i \phi}$ left after the transformation holds the information about the change in the Operator Space basis. \\

From the developments presented in the paper, suggest not only the validity of the HTA, but the powerful utilization of Operator Space Manifold Theory, which the HTA is apart of. It provides a unique toolbox to solve equations and represent formulations in Quantum Phase Space not seen elsewhere. 

\bigskip

\section{Future Developments and Acknowledgements}
\paragraph{Future Developments}
The main path of this work for the future is to further develop the Operator Space Manifold Theory to provide solutions to the Quantum Phase Space Dirac Equation. We also plan to use the developments seen here to develop configuration space formulations in the Quantum Phase Space.

\paragraph{Acknowledgements}
I would like to express my gratitude to the Gatton Academy and the Department of Physics and Astronomy at Western Kentucky University for their support in this research.

\bigskip

\section*{Statements and Declarations}
The authors declare that no funds, grants, or other support were received during the preparation of this manuscript. \\

Financial Interests: The authors have no relevant financial or non-financial interests to disclose. \\

The independent author, Gabriel Nowaskie, did all work concerning this paper from the writing of the manuscript to the calculations and data analysis. All authors read and approved of this final manuscript.

\bigskip 

\begin{thebibliography}{99}
\bibitem{a}
Torres-Vega, G.; Frederick, J.H.,
\emph{A quantum-mechanical representation in phase space.},
\emph{J. Chem. Phys.} {\bf 98} (1993) pg.3103-3120 \quad https://doi.org/10.1063/1.464085

\bibitem{b}
Simpao, V. A., 
\emph{Real wavefunction from Generalized Hamiltonian Schrodinger Equation in quantum phase space via HOA (Heaviside Operational Ansatz): Exact analytical results.},
\emph{J. of Math. Chem.} {\bf 52} (2014) pg.1136-1155 \quad https://doi.org/10.1007/s10910-014-0332-2

\bibitem{c}
Simpao, V. A., 
\emph{Toward chemical applications of Heaviside operational Ansatz: exact solution of radial
Schrodinger equation for nonrelativistic N-particle system with pairwise 1/rij radial potential in quantum
phase space.
},
\emph{J. Math. Chem.} {\bf 45} (2009) pg.129-140 \quad https://doi.org/10.1007/s10910-008-9372-9

\bibitem{d}
Nowaskie, G., 
\emph{The Half Transform Ansatz: Quarkonium Dynamics in Quantum Phase Space.} 
 (Version 2.0), \emph{ArXiv} (2023)\quad https://arxiv.org/abs/2303.16356



\end{thebibliography}
\end{document}